\documentclass[aps, prd, twocolumn, lengthcheck, superscriptaddress, showpacs, letterpaper, nofootinbib]{revtex4-1}

\usepackage{graphicx}
\usepackage{color}
\def\L{{\cal L}}\def\O{{\cal O}}

\def\la{\langle}\def\ra{\rangle}
\def\be{\begin{eqnarray}}\def\ee{\end{eqnarray}}
\def\lsim{\mathrel{\rlap{\lower3pt\hbox{\hskip1pt$\sim$}}
     \raise1pt\hbox{$<$}}} %less than or approx. symbol
\def\gsim{\mathrel{\rlap{\lower3pt\hbox{\hskip1pt$\sim$}}
     \raise1pt\hbox{$>$}}} %greater than or approx. symbol
\def\le{ \begin{array}{ll}}\def\re{\end{array}}

\def\lear{ \left( \begin{array}{cc}}\def\rear{\end{array} \right)}

\def\le{ \left( \begin{array}{cc}}\def\re{\end{array} \right)}

\def\bi{\bibitem}
\def\del{\partial}

\def\L{{\cal L}}\def\O{{\cal O}}
\def\la{\langle}\def\ra{\rangle}
\def\be{\begin{eqnarray}}\def\ee{\end{eqnarray}}
\def\lsim{\mathrel{\rlap{\lower3pt\hbox{\hskip1pt$\sim$}}
     \raise1pt\hbox{$<$}}} %less than or approx. symbol
\def\gsim{\mathrel{\rlap{\lower3pt\hbox{\hskip1pt$\sim$}}
     \raise1pt\hbox{$>$}}} %greater than or approx. symbol

\begin{document}

\title{``Pseudo-Conformal"  Sound Speed  in the Core of  Compact Stars}

\author{Mannque Rho}

\affiliation{
Univerisit\'e Paris-Saclay, Institut de Physique Th\'eorique,  CNRS, 91191 Gif-sur-Yvette c\'edex, France
}

\date{\today}

\begin{abstract}
We present the argument that ``pseudo-conformal" symmetry permeates from low density near nuclear matter to high density in the core of massive neutron stars. As a support of this argument, we describe how the quenched $g_A\approx 1$ in nuclei and the sound speed $v_s^2/c^2\approx 1/3$ in compact stars are controlled by emerging scale invariance in nuclear interactions. In our description, quasi-baryons could ``masquerade" de-confined quarks in the interior of compact stars.
 \end{abstract}

\pacs{}

\maketitle

%%%%%%%%%%%%%%%%%%%%%%%%%%%%%%%%%%%%%%%%%%%%%%%%%%%%%

%{ Won-Gi Paeng$^a$\footnote{\sf e-mail: wgpaeng@ibs.re.kr},
%Thomas T. S. Kuo$^b$\footnote{\sf e-mail: kuo@tonic.physics.sunysb.edu },
%Hyun Kyu Lee$^c$\footnote{\sf e-mail: hyunkyu@hanyang.ac.kr},\\
%Yong-Liang Ma$^d$\footnote{\sf e-mail: yongliangma@jlu.edu.cn} and
%Mannque Rho$^e$\footnote{\sf e-mail: mannque.rho@cea.fr}}
%\vskip 0.3cm
%{\em $^a$Rare Isotope Science Project, Institute for Basic Science, Daejeon 305-811, Korea}{\em $^b$Department of Physics and Astronomy, Stony Brook University, Stony Brook, New York 11794, USA }{\em $^c$Department of Physics, Hanyang University, Seoul 133-791, Korea }{\em $^d$Center of Theoretical Physics and College of Physics, Jilin University, Changchun, 130012, China }{\em $^e$Institut de Physique Th\'eorique, CEA Saclay, 91191 Gif-sur-Yvette c\'edex, France }

\section{The Approach: G$n$EFT}
It was found in \cite{PKLMR} that when two hidden symmetries presumed to be encoded in QCD, one scale symmetry and the other flavor local symmetry, were suitably incorporated into chiral symmetric effective field theory to capture low-energy nuclear dynamics and extended to a chiral-scale symmetric EFT -- coined as G$n$EFT in \cite{GnEFT}  --  to address the equation of state of compact-star matter, the sound speed of massive stars came out  precociously close to the conformal speed $v_s^2/c^2=1/3$ in the interior of the stars at a density $n\sim (3 - 5) n_0$. The trace of the energy momentum tensor (TEMT) is found to be not zero (even in the chiral limit) at that density, so the speed cannot be truly conformal. It was therefore referred to as ``pseudo-conformal." In this article, we explain how this surprising prediction could account for the possible presence  of ``deconfined quarks" in the core  of compact stars updating recent developments confronting microscopic approaches anchored on QCD proper being discussed in the current literature~\cite{Annala,Larry1,Larry2,pQCD}.  It turns out that the notion of  pseudo-conformality can  figure equally importantly in nuclear physics at low density. In Appendix, we show how the long-standing mystery of the quenched $g_A\simeq 1$ in Gamow-Teller transitions in nuclei can be resolved in G$n$EFT.

Given that the nuclear dynamics involved at non-asymptotic densities is intrinsically nonperturbative from the QCD point of view, the approach is inevitably anchored on effective field theory. 

The hidden symmetries, suitably incorporated, involve ``heavy degrees of freedom (HdFs)" that encompass the density range  from the regime of normal nuclear matter density $n_0\approx 0.16$ fm$^{-3}$, where standard chiral EFT applies,  to $\sim 7 n_0$ where it is most likely broken down.  In the framework of G$n$EFT that we have formulated, the hidden local symmetry is represented by the light-quark vector mesons $\cal{V}=(\rho,\omega)$ ``conjectured" to be (Seiberg-)dual to the gluons~\cite{komargodski}.  It turns out to be significant that this local symmetry is ``hidden" in the sense defined in \cite{HLS,suzuki}: They are to emerge from pion clouds as composite gauge bosons. For this phenomenon to take place, it is indispensable that there be the ``vector manifestation (VM)"~\cite{HY:PR} at some high density with the vector mesons  becoming massless. Where the VM density $n_{\rm VM}$ is located cannot be calculated in the effective theory. It turns out, however, that  the sound velocity behaves  qualitatively differently depending on whether   $n_{\rm VM}$ is near or asymptotically higher than the density of the star, .  

Our approach relies closely on that the scale symmetry enters via the ``genuine dilaton (GD)" formulated by Crewther and Tunstall~\cite{CT,crawling,GD}. The GD is associated with an infrared (IR) fixed point at which the QCD $\beta$ function vanishes, $\beta (\alpha_{\rm IR})=0$ (where $\alpha_s$ is the strong gauge coupling), realized in the Nambu-Goldstone mode with massless pions $\pi$ and dilaton (that we denote as $\sigma_{d}$)  with nonzero pion decay constant $f_\pi$ and dilaton decay constant,  $f_{\sigma_d}$, $f_\pi=f_{\sigma_d} \neq 0$. Most significantly for us, it  supports massive matter fields at the IR fixed point. A somewhat similar idea was put forward in \cite{DDZ}. The difference, if any, between these two ideas, we believe, would not affect our arguments. 

It should be stressed that this GD IR structure appears to be drastically different ~\cite{GD,DDZ}  from the IR structure of dilatonic Higgs models, that purport to go beyond the Standard Model. It is this basic difference that leads us to the pseudo-conformality, not the conformality~\cite{Larry1} -- which in our picture should set in at much higher density than in compact stars --  that  prevails in nuclear dynamics in density from $\sim n_0$ to the densities relevant to compact stars and possibly beyond before the dilaton limit, defined below, sets in~\cite{MR-reviews}. It should be remarked, however, the ``conformal dilaton phase" of \cite{DDZ} which borders between the conformal window and QCD could support the conformality discussed in \cite{Larry1}. How the QCD phase moves into the conformal phase, if present,  is not clear. 

The scale symmetry is incorporated into the HLS Lagrangian $\L_{\rm HLS}$~\cite{HLS} with the ``conformal compensator" field $\chi$~\cite{coleman} 
\be
\chi=f_\chi e^{\sigma_d/f_\chi}
\ee
which transforms linearly both in mass and length scales. (Note that $\sigma_d$, like the pion in chiral symmetry transforms nonlinearly under scaling.)  One can formally construct scale-invariant Lagrangian from $\L_{ \rm HLS}$ by multiplying a suitable power of the CC field $\chi$ such that the action is scale-invariant. The resulting scale-chiral Lagrangian can be written as
\be
\L_{\chi{\rm HLS}}=\L_{\rm inv}+V
\ee
where {\it all} scale symmetry breaking, including quark mass terms, are included in the ``dilaton potential" $V$.  One can rewrite this in terms of the power counting in both chiral and scale symmetries with the trace anomaly taken into account~\cite{CT,LMR} by implementing the standard power-counting in chiral symmetry~\cite{vankolck} with  the departure from the IR fixed point taken as
\be
\Delta \alpha_s\sim \O(p^2)\sim \O(\del^2).
\ee
The resulting scale-chiral expansion~\cite{LMR} -- that we call $\chi$PT$_{\sigma_d}$ expansion -- is a lot more  complicated than the standard chiral perturbation theory (SchiPT), but to the leading order in the scale-chiral expansion to which we will restrict, it turns out to be  quite simple.  

Next the  baryon fields $\psi^\dagger=(p\  n)$ are coupled into $\L_{\chi{\rm HLS}}$\footnote{We will be dealing with $N_f=2$ although the GD approach of \cite{CT} is for $N_f=3$.}\footnote{We won't go into details which can be found in the reviews~\cite{MR-reviews}. We briefly mention here that the HdFs brought in by the hidden symmetries are to play the role in G$n$EFT the putative ``hadron-quark continuity" that plays the crucial role in going, in what we consider to be hadronic variables, from nuclear matter at low density to compact-star matter at high densities.}. We call the resulting Lagrangian  $\L_{\psi \chi\rm HLS}$ with which G$n$EFT is formulated as we will explain.  In the scale-chiral power counting involving the vector mesons~\cite{HY:PR} and the scalar $\chi$~\cite{LMR},  this Lagrangian will then be of $\O(\del^n)\sim \O(p^n)$ with $n=1, 2, ...$. We will be mainly working with the leading order (LO) $n\leq 2$. 

Now instead of doing the systematic power expansion, highly successful in SchiPT, to high order in $\chi$PT$_{\sigma_d}$ expansion, which is in principle doable but cumbersome at best, we will instead develop an EFT that maps the resulting LO Lagrangian denoted  $\L^{\rm LO}_{\psi\chi{\rm HLS}}$ to a density functional theory (DFT) built on renormalization-group (RG) treatment of baryons on Fermi sea along the line developed in condensed matter physics~\cite{FL}. We apply the approach of how to go from chiral Lagrangian (in LO with suitable BR-scaling taken into account) to the ``Fermi-liquid fixed point  (FLFP)" theory, formulated in \cite{FR},  to the Lagrangian $\L^{\rm LO}_{\psi\chi{\rm HLS}}$. This is the principal tool in G$n$EFT that we will employ. Going beyond the FLFP can also be and will be done in ``ring approximations" in $V_{lowk}$ RG as described in \cite{PKLMR}. What is needed to access the compact-star densities in this G$n$EFT approach, not worked out in \cite{FR},  is how the HdFs effectuate the putative hadron-quark continuity (HQC) at $n \gsim 2 n_0$ in the Landau (fixed-point) parameters extracted from the Lagrangian.

In this approach, there is only one unique -- not a hybrid -- Lagrangian, an effective Landau(-Migdal) Lagrangian, that is to be valid over the whole range of densities involved\footnote{Migdal reformulated  Landau theory ior nuclear processes~\cite{migdal}. From here on, by ``Landau," we mean ``Landau-Migdal." We should point out that the Fermi-liquid structure described here focuses on bulk properties of matter, in particular, the equation of state (EoS), whereas in condensed matter physics, what is currently of high interest in EFT of Fermi liquid is  gapless Fermi-surface fluctuations, potentially leading to non-Fermi-liquid states~\cite{sonetal}.} with the fixed-point quasi-particle (QP)  mass or Landau mass $m_L$, two-body QP-QP interaction parameters $F$, $G$ etc. built from $\L^{\rm LO}_{\psi\chi{\rm HLS}}$.  The key points developed in \cite{PKLMR} (refined in \cite{MR-reviews} with errors corrected) are as follows:
\begin{enumerate}
\item Up to the density $n_0$, the parameters of the (Landau-Migdal) Lagrangian are controlled by the known properties of normal nuclear matter. Treated at the mean-field level, which corresponds to what's known as the Landau Fermi-liquid fixed point (FLFP) approximation (with $\bar{N}=k_F/(\Lambda_{\rm FS} -k_F)\to \infty$)~\cite{FL},  it  reproduces more or less all global properties of nuclear matter that are reliably described,  to N$^3$LO,  by SchiEFT. How the FLFP approximation fares in nature is aptly illustrated by the  anomalous proton orbital gyromagnetic ratio $\delta g_l^p$, the quenched $g_A$ in nuclei and enhanced axial-charge transitions in heavy nuclei~\cite{FR}. 
\item  The effect of the putative hadron-quark (HQ) continuity is brought into the Lagrangian by what is given by the topology change at $n\gsim 2 n_0$ from skyrmions to half-skyrmions when the topological baryons of Lagrangian $\L_{\chi{\rm HLS}}$ are put on crystal lattice~\cite{PKLMR}. Putting skyrmions on crystal lattice to describe baryon matter is evidently a very poor procedure at low densities (e.g., at $\sim n_0$ which is in Fermi liquid) but it is justified at high density and in the large $N_c$ limit which underlies the Landau Fermi-liquid effective field theory.  The precise densities involved cannot be pinned down in the theory. But the transition involves robust features topology brings in. The most crucial feature for us is that at the HQ changeover, the condensate of the bilinear quark fields $(q^\dagger q)$, while non-zero locally and supporting chiral waves, goes to zero when space-averaged at a density $n=n_{1/2} \gsim 2n_0$.  But the pion condensate $f_\pi$ remains nonzero, thus the quark condensate is not an order parameter of the chiral phase transition. It is somewhat like the ``pseudo-gap" phenomenon  in condensed matter physics~\cite{pseudogap}.
\item There arise various, highly remarkable, consequences when this topology change is incorporated into the Lagrangian  $\L_{\psi \chi\rm HLS}$ and treated at the mean field (FLFP) approximation. 

First:  There is a cusp in the nuclear symmetry energy $E_{\rm sym}$ wrapped by the $\rho$-meson cloud at the skyrmion-to-half-skyrmion density $n_{1/2}\gsim 2 n_0$. This brings in the soft-to-hard changeover in the EoS at that cross-over density~\cite{cusp} that naturally accounts for $K_0$, $E_{\rm sym} (n_0)$ and $L=3n \frac{dE_{\rm sym}}{dn}|_{n=n_0} $ to be consistent with nature\footnote{According to the recent summary, e.g., \cite{bal},  $K_0\approx 240$ MeV, $E_{\rm sym} (n_0)\approx 31.7$ MeV, $L(n_0)=57.7\pm 19$ MeV.}. 

Second: The coupling between the dilaton $\chi$ and the $\omega$, both playing a crucial role in the in-medium nucleon mass,  leads, in the chiral limit,  to the extremely simple result for the TEMT for $n\gsim n_{1/2}$~\cite{PKLMR}
\be
\la\theta_\mu^\mu\ra=\epsilon-3P=4V_d(\la\chi\ra) -\la\chi\ra\frac{\del V_d(\chi)}{\del\chi}|_{\chi=\la\chi\ra}.\label{temt}
\ee
In this formula the dilaton potential $V_d$  contains in addition the baryon fields. In doing this calculation, it is imperative that the density dependence of the parameters inherited from matching with QCD (via relevant current correlators and the VEV with the dilaton field) be treated in accordance with the thermodynamic consistency~\cite{thermodynamic}. Unless this is done correctly, one fails to arrive at (\ref{temt})~\cite{PKLMR}.

Third: $\la\chi\ra$ becomes density-independent for $n\gsim n_{1/2}$  only for high VM  density, $n_{\rm VM}\gsim  25 n_0$, and $\la\chi\ra\to c m_0$ (with $c$ a constant) where $m_0$ is the chiral invariant nucleon mass. This result, due to a close interplay -- with the $\rho$ decoupled from the nucleons -- between the dilaton $\chi$ and the $\omega$, signals the emergence of parity doubling in the baryon spectrum~\cite{interplay}. In our approach, the parity doubling is not put in {\t ab initio} as is done in the literature. It emerges from the interactions. It has been verified to remain valid when $\O(1/\bar{N})$ corrections to the FLFP approximation are made in the  $V_{lowk}$ formalism. That the TEMT (\ref{temt}) becomes density-independent for $n\gsim n_{1/2}$ is the key ingredient of the pseudo-conformality.
\end{enumerate}

\section{Predictions on compact stars}
We now turn to the predictions on compact stars given in  the theory G$n$EFT.

First we look at the density dependence of the TEMT. One gets a qualitatively correct answer by looking at the mean field (a.k.a., FLFP) approximation result (\ref{temt}). Taking the derivative with respect to density we have
\be
\frac{\partial}{\partial n} \la\theta_\mu^\mu\ra=\frac{\partial \epsilon (n)}{\partial n} \Big(1-3\frac{v_s^2 (n)}{c^2}\Big)\label{derivTEMT}
\ee
where the sound $v_s$ is inserted using $v_s^2/c^2=\frac{\partial P(n)}{\partial n}/\frac{\partial \epsilon(n)}{\partial n}$. 
Since the condensate $\la\chi\ra$ changes as density changes for the density regime $n\lsim n_{1/2}$, with $n_{1/2}$ being the topology change density, the TEMT changes with density. But as noted, the dilaton condensate tends to a density-independent constant $\propto m_0$, as $n$ goes above $n_{1/2}$, the right-hand side of (\ref{derivTEMT})  will go to zero
\be
\frac{\partial \epsilon (n)}{\partial n} \Big(1-3\frac{v_s^2 (n)}{c^2}\Big)\to 0 \ {\rm for}\ n\to n_{1/2}.
\ee
There is no reason to expect $\frac{\partial \epsilon (n)}{\partial n}\to 0$ (e.g., no Lee-Wick state), hence we arrive at what we call pseudo-conformal speed
\be
v^2_{\rm pcs}/c^2\to 1/3.
\ee
One arrives at the same  result in the $V_{lowk}$ calculation going beyond the FLFP approximation~\cite{PKLMR}. What's  found was that with the HQ changeover from below to above $n_{1/2}$, {\it strong nuclear correlations} intervene in the way the TEMT varies over the changeover density, with a ``huge" bump produced with the speed going toward the causality limit $v_s/c=1$\footnote{This becomes more prominent if the changeover density is $n\gsim 4 n_0$. In fact at this $n_{1/2}$ even the causality is violated~\cite{MR-reviews}.} and rapidly converging at increasing density close to, though not necessarily on top of, $v^2_s/c^2=1/3$~\cite{PKLMR}. The mechanism for producing such a big bump may be indicative of how the changeover from hadronic degrees of freedom in EFT to quark-gluon degrees of freedom in nonperturbative in QCD may be signaling the complexity involved in that region\footnote{This resembles how the HLS degrees of freedom (via the homogeneous or hidden Wess-Zumion term) wraps the cusp structure for the $\eta^\prime$ potential associated with the $U_A(1)$ anomaly~\cite{karasik}.}.

\begin{figure}[h]
\begin{center}
\includegraphics[height=5.2cm]{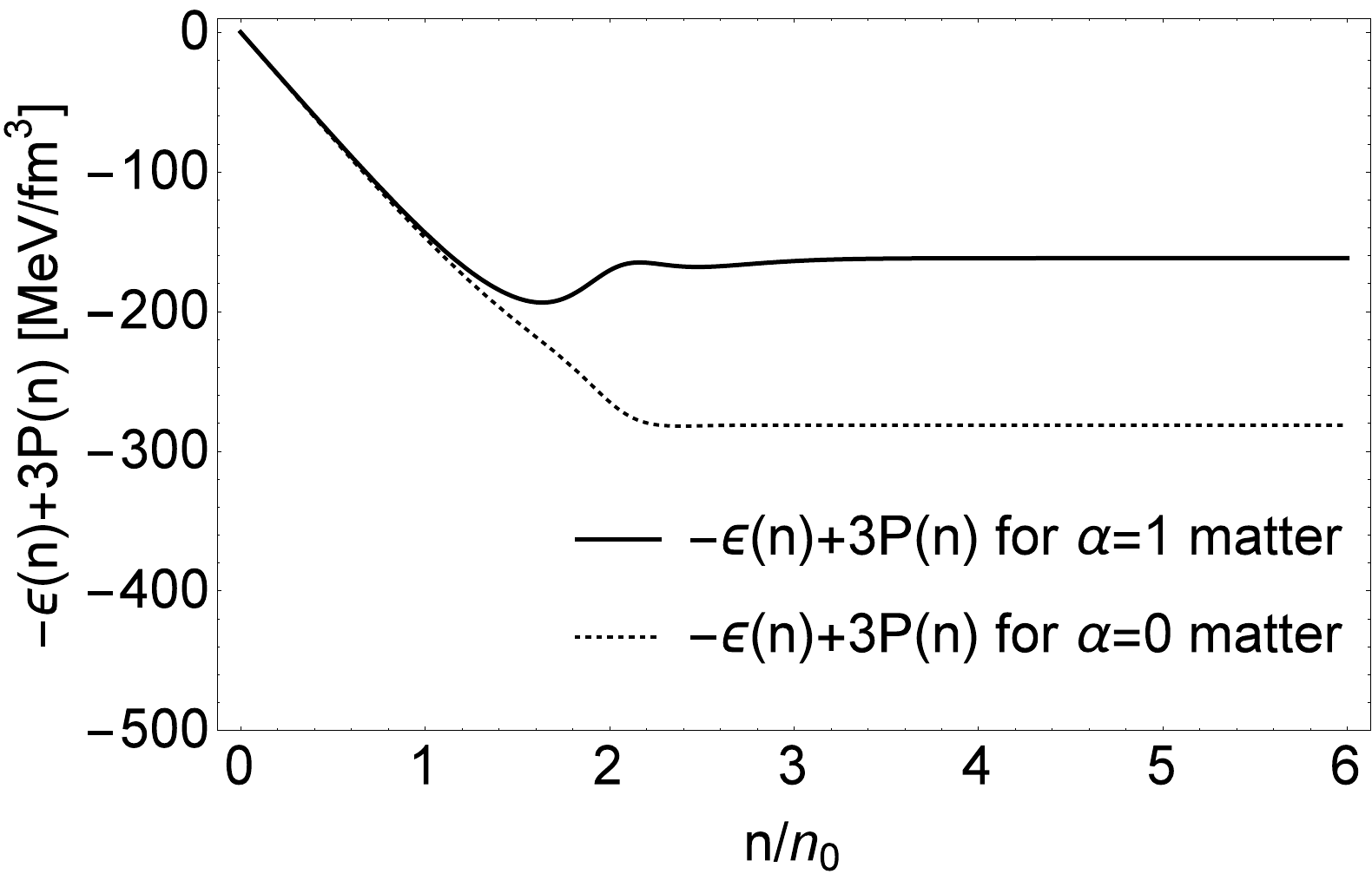}
\includegraphics[height=5.2cm]{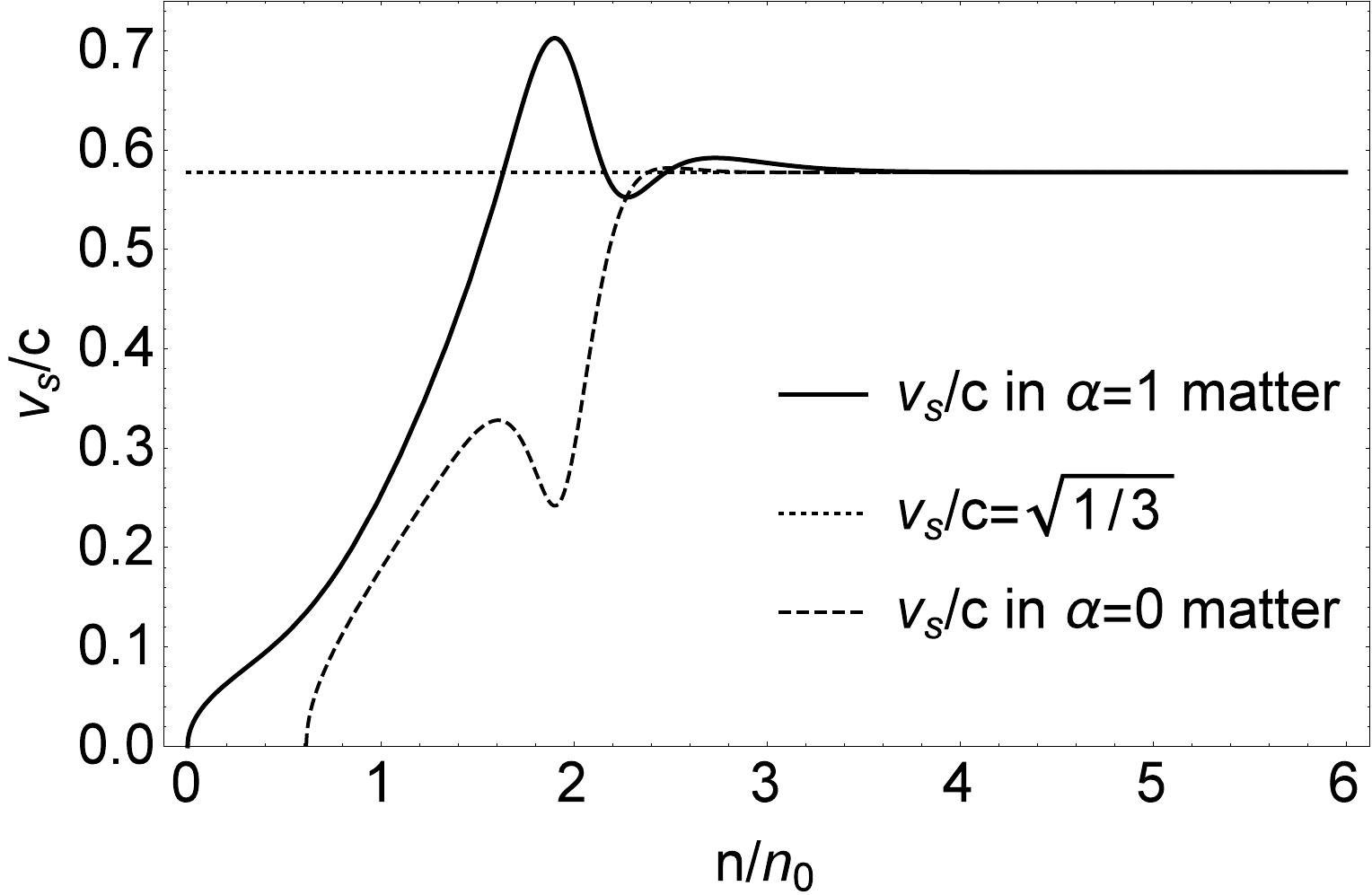}
\caption{ $-\epsilon + 3P$ (upper panel) and sound speed vs. density (lower panel) for $\alpha =\frac{N-P}{N+P}$ where $N(P)$ is the number of neutrons (protons).
 }\label{sound}
 \end{center}
\end{figure}

\begin{figure}[htbp]
\begin{center}
%\vskip 0.3cm
%\includegraphics[width=0.32\textwidth]{EoS.pdf}
\includegraphics[width=7cm]{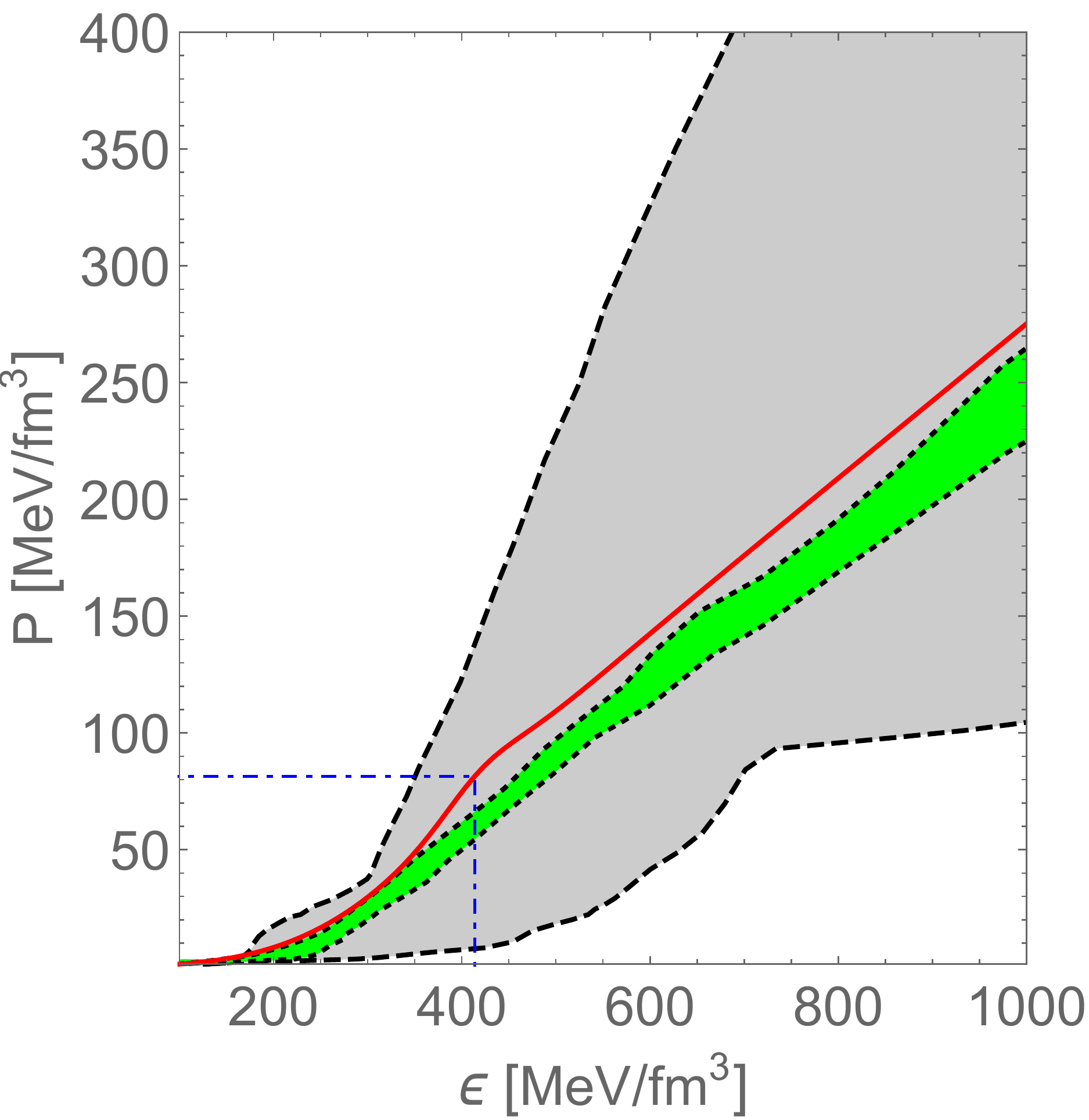}
%\vskip 0.5cm
\caption{Comparison of $(P/\epsilon)$  between the prediction and that generated with the
sound velocity (SV) interpolation method used in~\cite{Annala}. The gray band is  from the causality and the green band from the conformality. The red line is the G$n$EFT prediction\cite{AAPPS}. The dash-dotted line indicates the location of the topology (a.k.a. hadron-quark) change.}
\label{annala}
\end{center}
\end{figure}

The detailed structure depends on the location of the changeover density. It turns out, however, to be qualitatively the same in the range phenomenology  places, say, at $2\lsim n_{1/2}/n_0 < 4$~\cite{MR-reviews}.  The typical structure is illustrated in Fig.~\ref{sound} for $n_{1/2}=2n_0$. 
It is important to note that  in the range of densities involved in compact stars
\be
\la\theta^\mu_\mu\ra > 0
\ee
which gives
\be
\Delta=\frac 13 -\frac{P}{\epsilon} > 0
\ee
going independently of density for $n\gsim n_{1/2}$. It can be seen in Fig.~\ref{annala} that it is parallel and very close, with small deviation, to the band generated with the ``sound velocity interpolation method " used in \cite{Annala}. 
It should of course be stressed that the approach to the pseudo-conformal speed $v^2_{\rm pcs}/c^2\approx 1/3$ is {\it not} conformal $v_s^2/c^2=1/3$.  As noted below, at a  higher density approaching what is called ``dilaton-limit fixed point"~\cite{DLFP} where $\la\chi\ra$ goes to zero,  the sound speed should approach the true conformal speed.  We believe that  the DLFP density should be close to  the VM fixed point $n_{\rm VM}\gsim 25 n_0$, way outside of the range of densities involved in the stars. It is difficult to relate directly what is described in this G$n$EFT with the analysis made in \cite{Larry1}. It is however tantalizing that the sound speed given in Fig.~2 in \cite{Larry1} resembles the pseudo-conformal sound speed (Fig.~\ref{sound}). It indicates  that we are dealing here with strongly coupled (pseudo-)conformal matter. 

\begin{figure}[htbp]
\begin{center}
\includegraphics[width=0.4\textwidth]{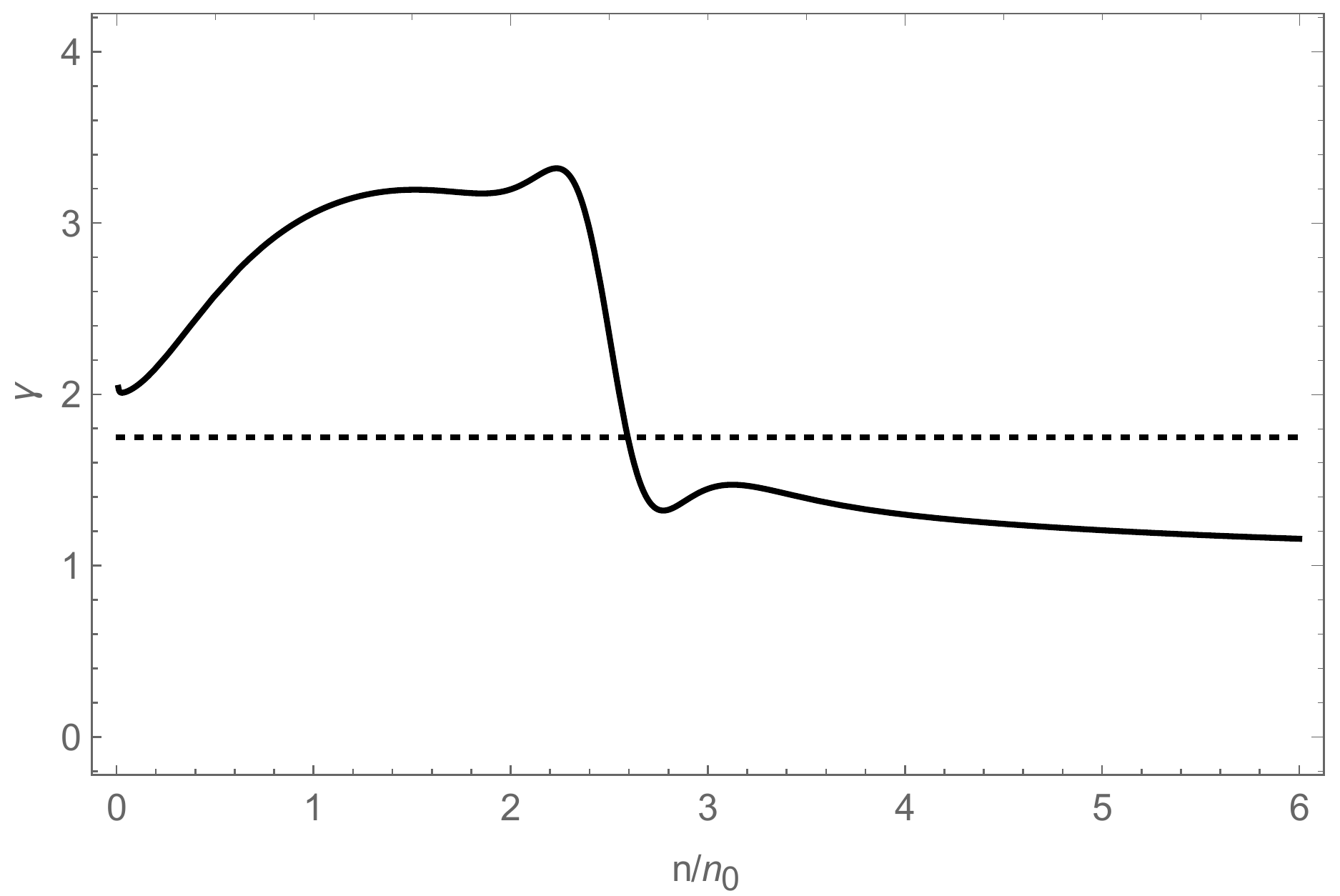}
\end{center}
\vskip -.5cm
\caption{Density dependence of  the polytropic index  in neutron matter. Here $n_{1/2}\approx 2.5 n_0$ was taken, so the changeover density -- strong correlation regions is shifted slightly upwards, but there is no qualitative difference.}
\label{polytropic}
\end{figure}

There are further indications that the pseudo-conformal matter resembles ``deconfined quark" matter. Let's look at  the polytropic index defined by
\be
\gamma= {d\ln P}/{d\ln \epsilon}.
\ee
Plotted in Fig.~\ref{polytropic}, it 
shows a large $\gamma\sim 3$ below $n_{1/2}$ expected in nuclear matter,  drops below 1.75 at the topology change and then goes to near 1 at the core density $\sim 6 n_0$ of the star. This reproduces what was identified as a signal  for ``deconfined" quark matter in  \cite{Annala}.  But there are basic differences between our system and what's described in \cite{Annala}. In our theory, conformality is broken, though perhaps only slightly at high density, in the system.  There can also be fluctuations around $v_{pcs}^2/c^2=1/3$ coming from the effects by the anomalous dimension $\beta^\prime$.  Thus the $v_{pcs}^2/c^2$ must fluctuate around 1/3, not on top of it.

\section{Consistency with pQCD top-down?}

The degrees of freedom that enter above the  hadron-quark continuity simulated by the topology change with what appear to  be fractionally-charged baryonic quasiparticles are not so outlandish if one imagines strong nuclear correlations in action as in the way electrons behave in strongly correlated condensed matter systems. In fact some {\it albeit} speculative but novel ideas along that line were recently discussed in \cite{symmetrypaper1,symmetrypaper2}. 

Now let's see  how these (somewhat speculative) ideas can mesh with what's predicted by QCD.

There have been tour-de-force efforts to arrive at the EoS relevant to the core of compact stars in perturbative QCD coming top down. Some of the results so far obtained appear to be quite relevant to what we have obtained in G$n$EFT. Among them are the results of the pQCD calculation given in \cite{pQCD} that render feasible even a semi-quantitative comparison with the G$n$EFT prediction. 

\begin{figure}[h]
\begin{center}
\includegraphics[width=7cm]{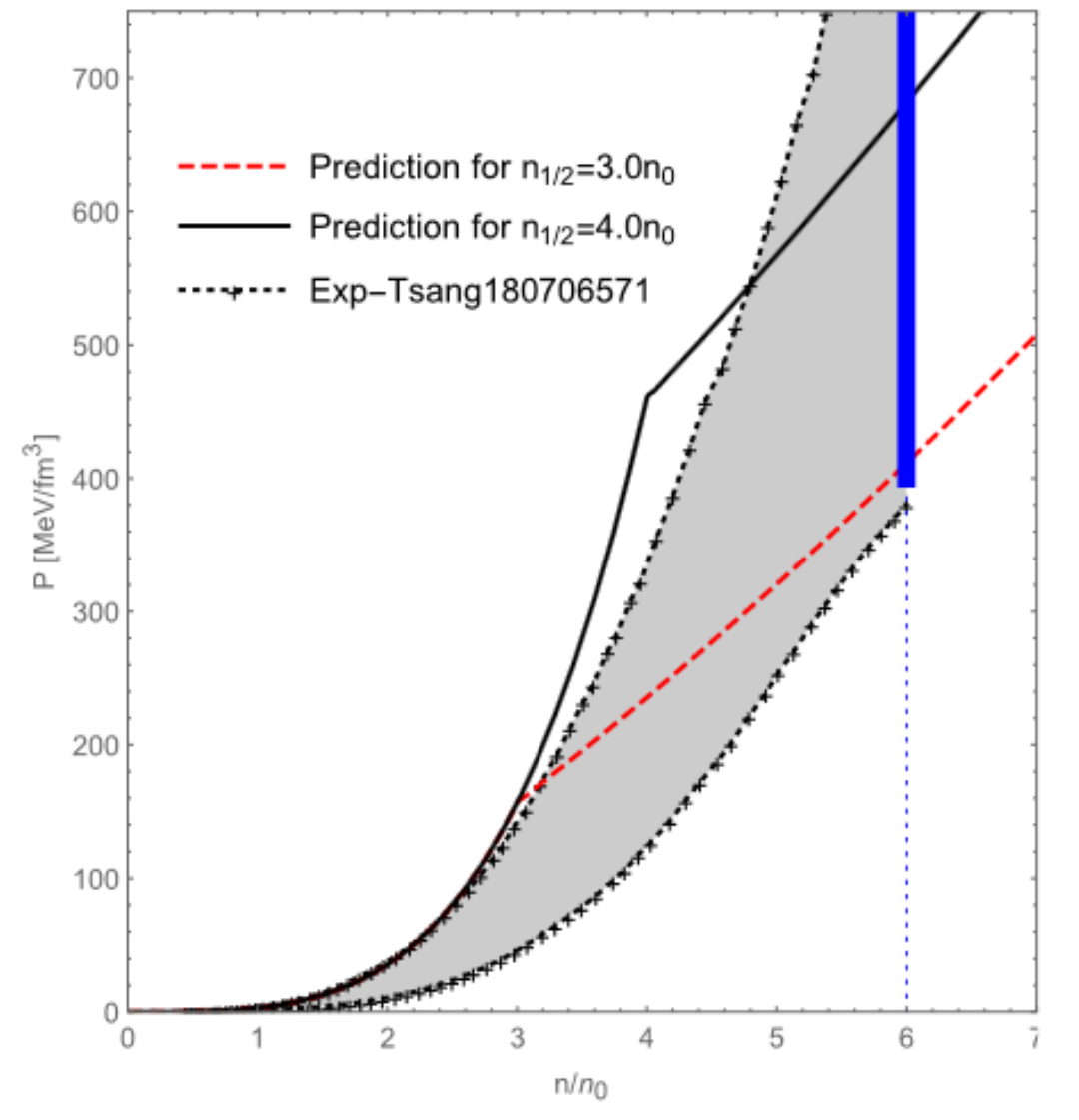}
\includegraphics[width=7cm]{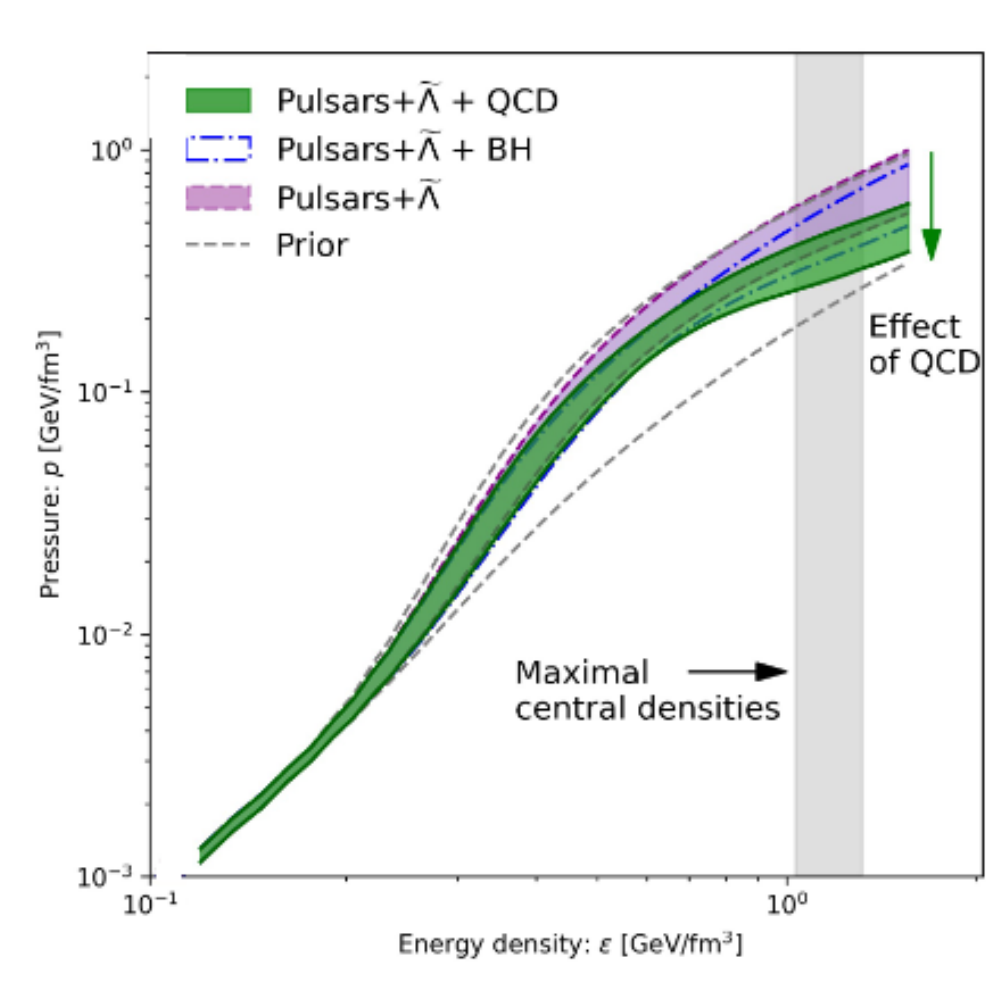}
\caption{Upper panel: Prediction in G$n$EFT for $n_{1/2}=3 n_0$~\cite{AAPPS}. $n_{1/2}=4n_0$ seems to be ruled out as the sound speed violates the unitarity bound $v_s^2/c^2=1$. Lower panel:  The state-of-the-art pQCD calculation~\cite{pQCD} indicting the effect of QCD. Note that the central density involved is $\sim 6 n_0$ in both cases.}\label{comp}
 \end{center}
\end{figure}

In Fig.~\ref{comp} (upper panel) is shown the G$n$EFT  prediction for the EoS for $n_{1/2}\approx 3n_0$, corresponding to the  hadron-quark crossover density.\footnote{Also shown is for $n_{1/2}=4n_0$ but it was ruled out in our scheme because it gives the sound speed violating the causality limit.} The prediction for $n_{1/2}=2n_0$ is slightly different from that of $n_{1/2}=3n_0$ but they both are equally trustful.  What's characteristic of this prediction is that the cusp in the symmetry energy $E_{\rm sym}$, ``wrapped" by $\rho$ meson cloud with dropping mass, is first hardened approaching  $n_{1/2}$\footnote{This makes the symmetry energy at $\sim 2 n_0$ bigger than what's given by standard chiral perturbation theory, then softened  before reaching the density of the star core at $\sim 6 n_0$ .  This feature is reflected in Fig.~\ref{comp} with $E_{\rm sym}$ bending after $n_{1/2}$ making the EoS softer toward the central density of the star.} and then softens as density increases beyond $n_{1/2}$.  It is easy to see how this softening behavior sets in in the Fermi-liquid fixed point approximation in G$n$EFT.  As noted, as one approaches the dilaton limit fixed point, the $\rho$ meson decouples from the baryonic matter (before the VM fixed point) and the remaining (heavy meson) degrees of freedom, the scalar (dilaton $\sigma_d$) and the isoscalar vector meson ($\omega$),  interplay to keep the nucleon mass go like $\sim m_0$. The quasiparticles invovoled are more or less free of interactions, hence preserving scale invariance (also seen in dense half-skyrmion simulation, Fig.~(11) (right panel) in \cite{PKLMR}). 

In Fig.~\ref{comp} (lower panel) is plotted the pQCD result of \cite{pQCD}. In \cite{pQCD}, the way QCD asymptotically high density calculations propagate down to lower densities subjected to ``thermodynamic consistency, stability and causality" is plotted. The calculation provides stringent  constraints, as indicated in the figure, to the EoS at densities relevant to the core of the star. 

The result clearly shows that pQCD softens the EoS of the most massive neutron stars going toward the central density of $\sim 6n_0$. Note that the stiffening and then softening in the EoS take place roughly at the same densities as in G$n$EFT. Precise matching would not make much sense given the approximations involved in both pQCD and G$n$EFT but there is a clear qualitative consistency.  It would be interesting to understand the stiffening-softening of the EoS in pQCD as in G$n$EFT where the interplay of scale symmetry and hidden local symmetry is found to play a crucial role in the emergent parity-doubling symmetry. 

\section{Conclusion with a bit of speculation}

In this brief review is recounted what has taken place, initiated in 2007 in the 5-year ``World-Class University Program" at Hanyang University in Seoul supported by the Korean Government, then continued at IBS (Korea) and at Jilin University in Changchun (China), with the objective of understanding ultradense matter stable against gravitational collapse. This issue has currently become a hot topic 
in physics with the advent of gravitational waves. The resulting approach with only hadronic degrees of freedom, implemented with hidden symmetries of QCD, with the parameters of the Lagrangian endowed with the vacuum sliding by density, G$n$EFT, is globally consistent with available data.

What is striking of this approach is that although no explicit QCD degrees of freedom are involved, the property of the core of massive stars predicted in this approach is uncannily similar to that of ``deconfined quarks" with one apparent difference on the nature of the sound speed.

How can this be?

We have no convincing  argument but one possible answer is this.

In skyrmion-half-skyrmion crystal simulations of dense matter, one can imagine how the half-skyrmions  can turn into fractionally charged baryons. Two half-skrymions are confined by a pair of monopoles, so they are not separable.  However,  surprisingly, the bound two-1/2-skyrmions are found to propagate scale-invariantly as observed in \cite{PKLMR}. Suppose  it is feasible to liberate the two half-skyrmions by suppressing the monopoles. Then it may be feasible to transform two half-skyrmions to three 1/3-charged objects~\cite{vento}.  One can then think of fractionally charged baryons populating the dense matter. Indeed in  condensed matter physics, with domain walls, there can be stacks of sheets containing deconfined fractionally charged objects behaving like ``deconfined quarks" coming from the bulk in which the objects are confined ~\cite{domain walls}.  In \cite{fraction}, highly speculative ideas along this line are entertained.

In conclusion, we propose that what's seen in the core of compact stars could be fractionalized quasiparticles masquerading ``deconfined quarks." As a test, we offer the ``duck test"~\cite{duck}: {\it If it looks like a duck, swims like a duck, and quacks like a duck, then it probably is a duck.}

\subsection*{Acknowldegments}

The author would like to acknowledge helpful  discussions on the matter treated in this review with, and suggestions on drafting this note from, Hyun Kyu Lee.

\setcounter{section}{0}
\renewcommand{\thesection}{\Alph{section}}
\setcounter{equation}{0}
\renewcommand{\theequation}{A.\arabic{equation}}
%%%%%%%%%%%%%%%%%%%%%%%%%%%%%%%%%%%%%%%%%%%%%%%%%%%%%%%%%%%%%%%%%%%%%%

\section*{Appendix: The case of quenched $g_A$ in light of scale symmetry}

A most surprising spin-off of the emergence of hidden scale symmetry in dense matter is that  the same pseudo-conformality also permeates at low density in certain channels of nuclear interactions. There has been a long-standing mystery that the axial-vector coupling constant $g_A$ in Gamow-Teller transitions in nuclei is observed to be quenched from the free-space value  $g_A=1.276$  to $g_A^{\rm obs}\approx 1$~\cite{gAreview}. This mystery which dates way back to early 1970's~\cite{wilkinson} could be given a simple resolution in terms of emergent scale symmetry in the same approach, G$n$EFT, applied at high density to compact stars.

In G$n$EFT, the nuclear axial current coupling to the EW current is scale-invariant -- modulo possible anomalous dimension corrections~\cite{MR-gA}
\be
J^{j}_{5\mu}=g_A\bar{\psi}\gamma\gamma_\mu \frac{\tau^j}{2} \psi
\ee
where $j$ is the isospin index and $\psi^\dagger=(p\ n)$ field. No conformal compensator field $\chi$ figures here in the construction of the Lagrangian.  In the mean field of $\L_{\psi \chi\rm HLS}$ (a.k.a. in the Fermi-liquid fixed-point approximation),  it turns out that one precisely reproduces  the same Landau-Migdal (LM) $g_A^{\rm LM}$ for the quasiparticle as was obtained in the simplified treatment made in \cite{FR},
\be
g_A^{\rm LM}=g_A q^{\rm LM} =g_A\big(1- \frac 13\Phi \tilde{F}_1^\pi)^{-2}. \label{LM}
\ee
Here $\Phi=f_\pi^\ast/f_\pi$ and $q^{\rm LM}$ is referred to in the literature as the ``$g_A$ quenching factor."  There are no corrections to it in G$n$EFT. The superscript $\ast$ stands for density dependence. $\Phi$ therefore represents the scaling of $f_\pi$ as the vacuum changes at varying densities. This quantity has been measured in deeply-bound pionic atoms up to $\sim n_0$~\cite{yamazaki}. Here $ \tilde{F}_1^\pi$ is the pion contribution to the Landau interaction parameter $F_l$, which is accurately calculable by standard chiral perturbation theory (SchiPT) for $n\lsim 2n_0$. The product $\Phi \tilde{F}_1^\pi$ in (\ref{LM}) turns out to be remarkably constant in the range $1/2 \lsim n/n_0 \lsim 1$, which covers the density range between light and heavy nuclei. As noted in \cite{FR}, (\ref{LM}) is effectively an in-medium Goldberger-Treiman relation anchored on low-energy theorem which is satisfied in the vacuum to a few \% accuracy and could be equally accurate in nuclear matter, although high-order SchiPT calculations, as far as we are aware, have not been performed yet. We will take (\ref{LM}) to be of the same accuracy.  

Remarkably the quenching factor (\ref{LM}) comes out,  numerically, to be
\be
q^{\rm LM}\approx 0.79 - 0.80\label{qf}
\ee
in the range of density involved -- with a few \% uncertainty -- in medium and heavy nuclei.
Thus the effective $g_A$ for the quasiparticle, $g_A^{qp\rm eff}$, is
\be
g_A^{qp\rm eff}\approx g_A^{\rm LM}\approx 1.0.
\ee

Now how this $g_A^{qp\rm eff}$ is related to the $g_A^{\rm obs}\approx 1$ has been a long-standing mystery in nuclear physics~\cite{gAreview}.

The formula (\ref{LM}) corresponds to the single-decimation procedure in Wilsonian renormalization-group flow calculation in the Landau fixed-point approximation~\cite{doubledecimation}. This means that the path integral is performed for the Gamow-Teller amplitude from the cut-off $\Lambda_{\rm FS}$ above the Fermi sea all the way to the top of the Fermi surface in the Fermi-liquid fixed point approximation (ignoring higher-order $1/\bar{N}$ terms in the $V_{lowk}$ RG). The {\it full} Gamow-Teller response  is given by the response of the single quasiparticle sitting on the Fermi surface multiplied by a constant that captures the {\it full} nuclear correlations
\be
M_{\rm GT}=q^{\rm LM} g_A \big(\sum_j (\tau^\pm_j \sigma_j)\big)_{Q_f Q_i}
\ee   
where $Q_{i,f}$ stands for single quasiparticle states. As defined,  it is $q^{\rm LM}$ that captures the {\it full} correlation effects. 
How this $q^{\rm LM}$ in the Fermi-liquid theory is related to the shell-model results is discussed in \cite{MR-gA}. (In short, in terms of high-order nuclear perturbation calculations in shell-model space, this procedure would be equivalent to connecting, with Goldstone diagrams,  the parent ground state to the daughter particle-hole state via a blob that contains all orders of particle-hole-bubble intermediate states with the intermediate energy going up to $\Delta E\sim 300$ MeV connected by the tensor forces.) Here we are assuming the quasiparticles involved are quasi-nucleons, ignoring resonances (e.g., $\Delta$s) etc. It is, in practice, difficult to identify the quasiparticle states $(Q_f Q_i)$ in Ferm-liquid theory with the corresponding shell-model states.  Furthermore the observation of $g_A^{\rm obs}\approx 1$ is in light nuclei and the Fermi-liquid approach adopted is, however, more likely applicable to nuclear matter, so how to access shell-model states needs to be addressed. 

%It seems feasible in light-nuclear systems but it is most likely more applicable in heavy nuclei as the treatment made in this paper, as in \cite{FR}, is for nuclear matter. 

Although one cannot make one-to-one correspondence between the shell model states and the Fermi liquid states, what appears to be the most suitable for mapping the FLFP result in Landau-Migdal Fermi-liquid theory to shell-model in heavy nuclei is the doubly magic closed shell nuclei such as $^{100}$Sn proposed in \cite{MR-gA}.  The situation is not quite clear how to experimentally zero-in on the lowest shell-model daughter state  in $^{100}$In that best corresponds to the quasi--particle-quasi--hole excitation on the Fermi surface. So there will remain certain uncertainty that requires to be scrutinized in analyzing experimental results. 
%What will be needed to ultimately clarify the issue are new precision re-measurements and careful analyses of the data on the supperallowed Gamow-Teller transition from $^{100}$Sn to $^{100}$In. 
One can, however, make an interesting observation as it stands  from the presently available data on this transition.

What's involved  is the strongly enhanced Gamow-Teller transition of a proton in the completely filled shell (i.e., $g_{9/2})$ in $^{100}$Sn  to a neutron in the empty shell (i.e., $g_{7/2}$) in $^{100}$In with zero momentum transfer and $\sim$ zero MeV energy transfer. This  involves a transition as close as possible in kinematics to the decay of a quasiproton to a quasineutron on the Fermi surface in the Landau-Migdal theory. The square of the Gamow-Teller transition matrix element going from the pure proton $g_{9/2}$ shell to the pure neutron $g_{7/2}$ shell, referred to as ``extreme single-particle shell-model (ESPSM)" strength, is given by
\be
{\cal{B}^{\rm ESPSM}_{\rm GT}}= (160/9). 
\ee  
Suppose we have the exact wave functions -- which are of course not yet available for the given parent and daughter states -- and hence have the exact Gamow-Teller strength for the transition with the given scale-invariant Gamow-Teller operator that we write $ \cal{B}^{\rm exact}_{\rm GT}$, the ``exact" quenching factor will then be
\be 
q_{\rm th}^{\rm exact}=\big({\cal{B}^{\rm exact}_{\rm GT}/\cal{B}^{\rm ESPSM}_{\rm GT}}\big)^{1/2}.
\ee
Now since we don't know the exact wave functions, we don't know what $\cal{B}^{\rm exact}_{\rm GT}$ in the shell-model is. But nature, i.e., an experimental value  accurately  measured, will give it. Call it  $\cal{B}^{\rm nature}_{\rm GT}$. Measuring the GT strength will give us what the quenching factor is in nature. {\it It is the prediction of G$n$EFT at the FLFP approximation  that  $ q_{\rm th}^{\rm exact}$ should be given by (\ref{qf}).} 

In order to see how this prediction fares with nature, we pick the measured transition strength to the single daughter state, i.e., neutron in $g_{7/2}$. While not rigorously established, the transition seems to go to this state more than 95\% in certain model calculations, so let's simply assume that it goes 100\% and take what's measured in \cite{GSI} corresponding to  ``nature" \`a la GSI  
\be
{\cal{B}_{\rm GT}^{\rm nature}} \approx 10. 
\ee
This would give the  measured quenching factor  -- within the uncertainty involved (say, $\sim 5\%$?),
\be
q^{\rm nature}\approx 0.75
\ee
hence the quenched $g_A$ is
\be
g_A^{\rm nature}\approx 0.96.
\ee
Thus we have, within the possible uncertainty involved, 
\be
 g_A^{\rm LM} \simeq g_A^{\rm nature}.
\ee

We should underline two points here: 

One is that $g_A^{\rm LM}$ represents the effect of {\it solely} nuclear correlations,  not  a {\it fundamental" renormalization} in the given EFT of the coupling $g_A$ as was suggested by some workers in the field.  The nuclear correlations as formulated contain not only the single-particle operator but also many-body (meson-exchange) currents  involving nucleon fields.  The constant $g_A^{\rm LM}$ is therefore not a renormalized constant in nuclear axial response functions that applies to non-supperallowed GT transitions such as axial-charge (i.e., first forbidden) transitions,  double-neutrino and neutrinoless Gamow-Teller transitions. 

The other is that $g_A^{\rm LM}\to 1$ in nuclear matter is an intricate consequence of hidden scale symmetry emerging in medium at low density.  

Now what can one say as density increases toward the IR fixed point density? To see what happens at high density
we reparametrize the scale-chiral field as
${\cal{Z}}=U (\chi/f_\chi)$ where $U=e^{i\pi/f_\pi}$ is the chiral field and then let ${\rm {Tr} (\cal{Z} \cal{Z}^\dagger)} \to 0$ in $\L_{\psi\chi{\rm HLS}}$ treated in the mean-field. We do this to drive the system to  the dilaton-limit fixed point density $n_{\rm dlfp}\sim n_{\rm VM}\gsim 25 n_0$. In order to prevent singularities from developing, one is constrained to impose the conditions~\cite{DLFP}
\be
g_A\to 1, \ f_\pi\to f_\chi.
\ee
Thus  the effective coupling constant for the quasiparticle, $g_A^{qp\rm eff}=1$, is seen to maintain the same value  from $n\sim n_0$ due to the nuclear correlations controlled by scale invariance to $n_{\rm dlfp}\sim n_{\rm VM}\gsim 25 n_0$ at high density as $m_{\sigma_d}\sim \la\chi\ra \to 0$. We interprete this as a manifestation of the pseudo-conformal symmetry permeating from finite nuclei at low density to high compact-star density and to even higher densities approaching $n_{\rm dlfp}\gsim 25 n_0$. 

\section*{Note added in proof}
The referees to this review raised several constructive issues in their comments. They do not directly address the thesis developed in this paper, but we find them most likely be relevant to the future development along the line of thinking adopted in this article.  We would like to respond to their comments to the best feasible.

There were three issues raised.

The first, raised by one of the referees, is the possible relevance of the color-flavor locked (CFL) state in three-flavor QCD at high density. The question is: Given that the CFL state is conjectured  to emerge at asymptotic density in QCD, what does the scenario developed in this article (anchored on the notion of hadron-quark (HQ) continuity) predict for the sound speed (SS) in the CFL phase? We cannot give a precise answer to this question in the framework of the pseudo-conformal notion, but it is very likely that the SS, $v_s^2/c^2$, is not conformal (that is $\neq  1/3$) in the CFL phase,  given that the trace of the energy-momentum tensor (TEMT) is not likely to be zero even at the asymptotic density.  This of course does not exclude the possibility of (so far unobserved) ``quark stars,” possessing drastically different equation of states (EoSs).

Briefly restated, the pseudo-conformal structure adopted in this review -- which is anchored on the putative HQ continuity (with no phase transitions) via the topology change mediated by the hadronic HdFs governed by hidden symmetries -- cannot be naively extended to too high a density. The highest density relevant in the given formalism is the vector manifestation density or the dilaton-limit fixed point $\gsim 25n_0$. What’s pertinent to the issue is the currently controversial question as to whether the Schaefer-Wilczek conjecture that underlies the HQ continuity at the asymptotic density is even valid. There are arguments~\cite{no-go-theorem}, though not generally agreed upon in the field, that it is indeed {\it invalid}, according to which ``the hadronic (e.g., nuclear) matter and the quark matter present must necessarily be separated by a phase transition as a function of density.”  

The second issue is the property of the compact-star SS deduced from a huge number of randomly generated EoSs that satisfy theoretical and observational constraints~\cite{ecker}. A number of non-trivial correlations in softness and stiffness of EoS at different density regimes reveals that the SS tends to decrease below the conformal limit in the core of the star, while it increases beyond conformal in outer layers. Although not ruled out by these results, the pseudo-conformal structure sound speed is found to deviate distinctively from them. It exhibits a maximum below the putative HQ transition density $n_{1/2}$, hence in the outer layer of the star, with the height dependent on the cross-over density, and converges {\it precociously} to  $v_s^2/c^2$ near 1/3 toward the star center. Within the approximations made, one expects certain degrees of fluctuations from 1/3 but not as wildly varying as seen in the agnostic analysis. It converges to $\sim 1/3$ toward the core. This is also predicted in the microscopic description of \cite{Larry1}, with the convergence pushed to higher densities.  It has been noted in \cite{cusp} that there seems little correlation, if any, between global star properties and the plethora of bump structure in the sound speed. Given that the cross-over region is the least well controlled theoretically, EFT bottom-up and QCD top-down in density, this bump structure may be hardest to resolve theoretically.

Finally, the third issue is what role the SS plays in the causal maximum mass limit of neutron stars in the framework of modified gravity such as e.g., $f(R)$~\cite{fR}. Such an issue addresses whether, or in what way, modified gravity can make impacts on the structure of the EoSs one extracts from the astrophysical observables. It appears that for a given EoS, the speed of sound can be considered even as a free variable.

In this third category of issues, there is one that addresses whether the physics of sound speed in massive compact stars can make impacts on a test of string theory in gravitational sector. In a post-Newtonian analysis of double field theory (DFT) as a test of string theory in gravitational sector, it has been suggested~\cite{choi-park} that “Eddington-Robertson-Schiff parameter” $\gamma_{\rm PPN}$ could be extracted from the interior structure of the proton. In a private communication~\cite{park}, J.-H. Park inquired to the author whether the vanishing of the TEMT in the core of massive compact stars discussed in the main text of this review could not be interpreted as the vanishing of  the TEMT in the proton interior and hence provide a signal for $\gamma_{\rm PPN}\simeq 1$. The possible conformal speed $v_s^2/c^2=1/3$~\cite{Annala,Larry1,Larry2} or the pseudo-conformal speed $v_{\rm pcs}^2/c^2\approx 1/3$ could perhaps be translated to a quantity corresponding  to $\gamma_{\rm PPN}\simeq 1$.

What seems tantalizing in the way the pseudo-conformal speed is formulated is that certain  notion, say, pseudo-conformality,  borrowed from condensed matter physics may overlap even with what’s at issue in gravity theory. 

\end{document}